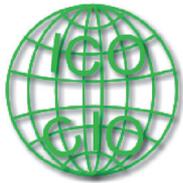

# NEWSLETTER

Commission Internationale d'Optique ● International Commission for Optics

# Getting used to quantum optics…

… and measuring one photon at both output ports of a beam splitter.

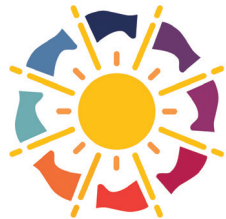

INTERNATIONAL YEAR OF LIGHT 2015

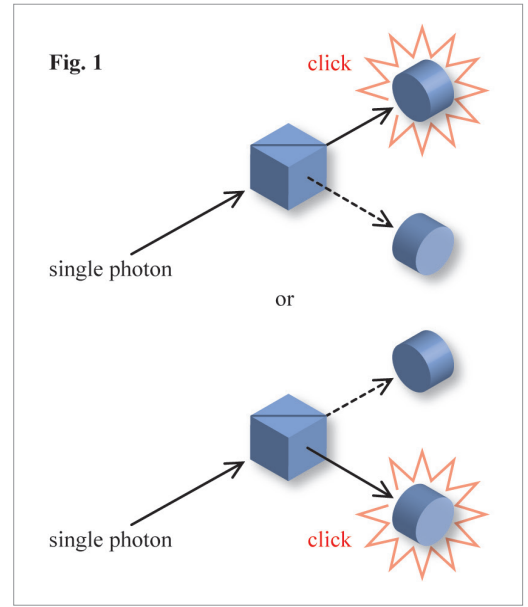

Fig. 1

single photon — click

or

single photon — click

When asked whether a photon can be split, one familiar with parametric down conversion (PDC) is tempted to answer yes, of course. In PDC, a photon is absorbed and two new ones are created, their energies adding up to the energy of the initial photon. In a way one might say that the PDC interaction is a perfect beam splitter, the two output beams being perfectly correlated. This nonlinear optical process is one of the workhorses in quantum optics and deserves many pages of appraisal[1] – but it is not the subject of this article. In this article I address the question of whether or not a single photon is split on a normal beam splitter, which is a linear optical element[2] – split in the sense that it will simultaneously affect the signal measured by two detectors in the two output ports of the beam splitter.

Let us begin by asking the fundamental question: What is a photon? Interestingly, this question (one to which there have been multiple, and sometimes conflicting, responses) takes us to Maxwell's equations, which celebrated their 150th anniversary in the International Year of Light, 2015. Maxwell wrote down his famous equations well before Fitzgerald first published what was later called the Lorentz transformation and well before quantum theory was formulated. Yet in a way, Maxwell anticipated the later developments: The equations in vacuum are invariant under Lorentz transformation, and their solutions are functions describing the modes in which the quantized field excitations live. And on top of this, it seems as if Maxwell's equations are also closely related to the properties of the quantum vacuum (see appendix).

Light as described by Maxwell's equations has four degrees of freedom (DOF), the helicity and the three components of the momentum vector. These can be translated into the polarization, the transverse mode profile (two DOFs) and frequency. The four degrees of freedom provide the space where quantum excitations "live". Since the quantum description of field modes is analogous and mathematically identical to that of a quantum harmonic oscillator, it is not surprising that the energy spectrum of the light field is comprised of equidistant energy levels, describing the energy in the mode. If the mode is in the $n$th energy eigenstate, one says that there are n photons in the mode. In that sense a single photon refers to the first excitation of this mode. For many practical purposes a mode can be thought of as an object with finite spatial extent and which may be moving with time. The first excited state of this moving mode is called a single photon wave packet, or simply a single photon.

In most systems of interest, multitudes of photons exist simultaneously. We thus ask the question, can we produce a single photon, i.e., is it possible to put a single quantum of energy into a particular mode of a system? The answer is yes, and interestingly, one standard method is by parametric down conversion. We do not know when we will detect a photon produced by PDC, but we do know that it is accompanied by a twin. Thus, upon detecting one photon of the pair we can be sure that the other one is there: we have a conditionally prepared single photon wave packet, conditioned on the detection of the twin.

---

[1] Theory: W H Louisell, A Yariv, A E Siegman, *Physical Review* 124, 1646 (1961); and D N Klyshko, *Soviet Phys. JETP Letters* 6, 23 (1967) Experiment: (3 in one year): S E Harris, M K Oshman, R L Byer, *Physical Review Letters* 18, 732 (1967); D Magde and H Mahr, *Physical Review Letters* 18, 905 (1967); and S A Akhmanov, V V Fadeev, R V Khohlov, O N Chunaev, *JETP Letters* 6, 85 (1967)

[2] A Luis, L L Sánchez-Soto, *Quantum and Semiclassical Optics* 7, 153 (1995)

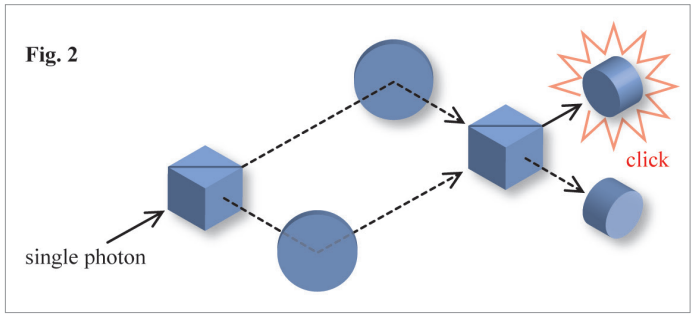

Fig. 2

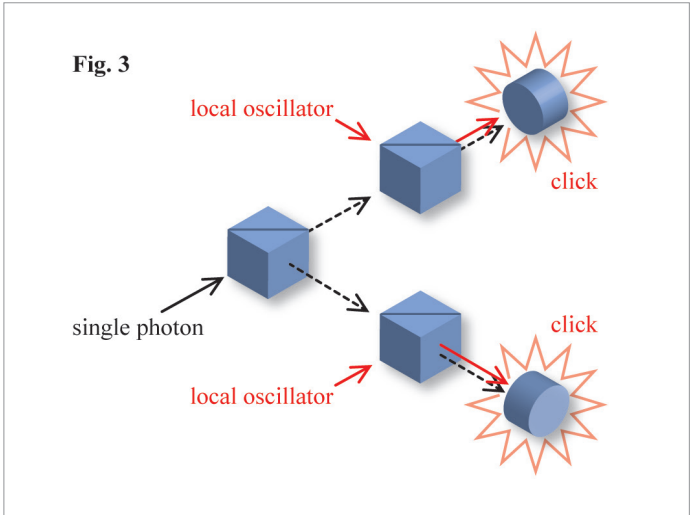

Fig. 3

Such a single photon wave packet we now let impinge on a beam splitter. Conventional wisdom tells us that while the beam splitter may well split a macroscopic beam of light, it does not split a single photon: a single photon is undividable (at least under linear interaction); it is either transmitted or reflected but not both. This statement we want to challenge.

First, we remind ourselves of the reason for this statement. In the standard arrangement a single photon is sent to a beam splitter at the outputs of which two click detectors are placed. These are detectors capable of detecting single photons. A short electrical pulse "click" indicates the absorption of a photon by the detector, as suggested in fig. 1. In such a scenario only one detector clicks in response to a single input photon[3,4,5]. Hence the statement that a photon is undividable and will be either reflected or transmitted. One can even formulate a model ascribing a stochastic property to the beam splitter responsible for sending the photon out of either of the two output ports. This is alright; the model describes the experiment satisfactorily. Yet, deep inside one may sense a little grumbling: is the beam splitter not a unitary element and would that not exclude any stochastic operation? And yes, there is this other view: the beam splitter "divides" the photon and sends it out both ports. In quantum physics this is described by the superposition of two quantum states, one photon in the first output port and none in the second super imposed with no photon in the first and one photon in the second output port[6]. In this model the detectors introduce the stochastic element. Whether or not they click is stochastic, with the boundary condition that the detection process "projects" the quantum state to a localized single photon state at the detector, the detection process requiring the full energy of the photon. The other detector will then never click in coincidence. If the experiment is repeated many times one finds that each detector has a 50% probability to click. This projection is the conceptually difficult part of quantum physics: in the quantum theoretical description an object is viewed as a wave described by a wave function, a concept well familiar from everyday life with interference, dispersion etc.; but in a quantum measurement much of the information content is lost, the object 'appearing' at one particular location (projection) although the wave function describing the object before detection was delocalized. The most common way to come to terms with this concept of 'projection' is to perform as many experiments as possible and to educate oneself to develop the proper understanding and intuition. So let us do the next experiment.

We now test whether the stochastic model of the beam splitter properly describes the more complex situation encountered when the two output beams of the first beam splitter are recombined on a second beam splitter. Obviously the system is an interferometer and the stochastic model does not work. For identical 50% beam splitters it would predict the photon to exit stochastically on either output port of the second beam splitter, independent of any change in the difference of the optical path lengths of the two interferometer arms. The contrary is what one observes. If the arm lengths are exactly equal the photon will exit from the symmetric[7] interferometer output, as illustrated in fig. 2. If the path length difference is changed, the count rates at the two output ports are complementary varying in a sinusoidal manner such as to add up to the constant entrance count rate. But this, one may argue, is not the decisive measurement, because we do not measure through which arm of the interferometer the photon travels. Two decades ago this question was open for a while.

Then, back in 1991, Tan, Walls and Collett[8] proposed repeating the single beam splitter experiment, this time not with click detectors measuring the energy in the beam but rather with detectors measuring electromagnetic fields, so called field quadratures, using

> A homodyne detector is comprised of a local oscillator, a beam splitter and a direct photo detector. Measurement by a homodyne detector projects the excitation of the mode onto a precise value of an electric field component (called field quadrature). The phase of the local oscillator determines which field quadrature is measured. Heisenberg's uncertainty principle requires that the eigenstates of the field quadratures have infinite energy. Therefore, in practice, one will always measure the field quadrature with a residual uncertainty. Measuring both output ports of the beam splitter inside the homodyne detection device and taking the difference of the two direct detector readings is called balanced homodyne detection. With field quadrature detectors one and the same "photon" can induce measurement results in two different detectors, unlike in the case of click detectors.

homodyne detectors. With homodyne detection there is no projection onto an energy eigenstate of the field. Of course, ultimately the photo-detector measures energy, but this energy is provided by the superposition of the single photon and the local oscillator[9]. Ideally, a homodyne detector rather projects onto a field quadrature eigenstate of the signal field. Thus, without any excitation in the signal mode, homodyne detection can be used to measure the zero point electric field uncertainty of the light[10], one application being a high-speed quantum random number generator[11]. For a related measurement of the field variance of the ground state see Riek et al[12].

Implementing the proposal by Tan et al[5] did not seem easy. So 10 years later, Björk, Jonsson and Sánchez-Soto[13] proposed a variation that combined a local oscillator with click detectors instead of the less sensitive direct photo detectors and that seemed more accessible experimentally. Then in 2003, Hessmo, Usachev, Heydari and Björk[14] actually conducted the latter experiment. A simplified illustration of the experimental system is shown in fig. 3. The experiment showed that a single photon at the input of the beam splitter caused correlated clicks at the two detectors. Depending on the relative phase of the two local oscillators the clicks were correlated, anti-correlated or uncorrelated. With the input photon blocked, there was no correlation for any relative local oscillator phase settings. This was the decisive experiment, which resolved the dispute on whether or not the superposition state displayed in footnote 3 qualifies as quantum entanglement[15]. The experiment is also conceptually quite important because it means that a single photon can have a simultaneous influence at two distant locations. The experiment proposed by Björk, Jonsson and Sánchez-Soto[9] and performed by Hessmo et al[11] is a seminal experiment that has not yet received the recognition it deserves.

In hindsight, another seminal paper by Lvovsky, Hansen, Aichele, Benson, Mlynek and Schiller[16] also involved the splitting of a single photon. In their experiment a single photon interfered with a local oscillator on a beam splitter and two direct detectors measured the signal at the two output ports comprising a homodyne detector. The difference of the signals measured with the two direct detectors (see box) was recorded and used as the input data for tomographic reconstruction of the single photon Wigner function[17], requiring the experiment to be repeated many times. The correlation between the signals at the output ports was, however, not recorded separately. Two such set-ups would have been required for the experiment proposed by Tan et al[5].

Closely related to the experiment by Hessmo et al.[10] and submitted and published only a few weeks later is the work by Babichev, Appel and Lvosky[18], who used homodyne detection following more closely the proposal by Tan et al.[5] By now several groups have exploited the simultaneous detection of a single photon with two detectors[19,20]. One may not only detect a single photon simultaneously at two different locations – with the help of local oscillators – but one can also measure a single photon at the same location at different times. Gulati, Srivathsan, Chng, Cerè, Matsukevich and Kurtsiefer[21] have performed such an experiment recently looking at the fluorescence of a single Rubidium atom using a local oscillator. The data shown are accumulated over many measurements, but in principle a single measurement on a single photon should show the exponential shape of the wave packet with a signal to noise ratio of two provided the whole solid emission angle is recorded. Collecting the fluorescence with a deep parabolic mirror should make this demonstration feasible[22].

As a conclusion one might say that nature challenges us if we want to cast it into models and theoretical frames. At times one recognizes that a concept that was helpful throughout most of our lives as physicists has to be modified. In the middle of this process of modification there may be heated discussions and disputes, and for a while one may be reminded of a phrase one can read in a paper by Elishakoff:[23] "The world is divided into people who think they are right". But the good thing about science is that in the end the dispute converges because scientists have learnt the survival strategy: if new experimental results challenge the old view – such as a single photon being either reflected or transmitted at a beam splitter – then one has to accept this, be flexible and develop a more appropriate point of view.

### Acknowledgement
It is my great pleasure to acknowledge helpful and enlightening discussions with Elisabeth Giacobino, Luis Lorenzo Sánchez-Soto, Maria Vladimirovna Chekhova, Markus Sondermann, Christoph Marquardt, William T Rhodes and Angela M Guzman.

### Appendix
One may speculate about a possible (and maybe obvious) connection between classical optics, i.e., Maxwell's equations, and the modern quantum vacuum, which is not void but filled with virtual elementary particles. A light field polarizes this vacuum and any linear response of the vacuum must be part of Maxwell's equations. In this sense Maxwell's displacement would be merely the sum of the polarization of the vacuum and real matter and the linear response of the vacuum is already accounted for in his equations: $\vec{D} = \varepsilon_0 \vec{E} + \vec{P} = \vec{P}_{vac} + \vec{P}_{mat.}$ See G Leuchs, A S Villar, L L Sánchez-Soto, *Appl. Phys. B* 100, 9 (2010) and G Leuchs, L L Sánchez-Soto, *Eur. Phys. J. D* 67, 57 (2013).

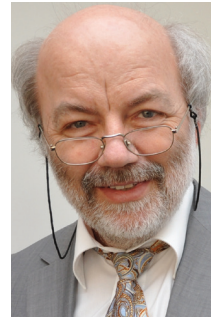

**Gerd Leuchs, professor of physics, Institute for Optics, Information and Photonics, Univ. Erlangen. Director of the Max Planck Institute for the Science of Light in Erlangen. His research spans the whole range from classical to quantum optics.**